# QuickLoc: Adaptive Deep-Learning for Fast Indoor Localization with Mobile Devices

Saideep Tiku

> Department of Electrical and Computer Engineering, Colorado State University, Fort Collins, Colorado, USA, saideep@colostate.edu

Prathmesh Kale

> Department of Electrical and Computer Engineering, Colorado State University, Fort Collins, Colorado, USA, prathmesh.kale@colostate.edu

Sudeep Pasricha

> Department of Electrical and Computer Engineering, Colorado State University, Fort Collins, Colorado, USA, sudeep@colostate.edu

## ABSTRACT

Indoor localization services are a crucial aspect for the realization of smart cyber-physical systems within cities of the future. Such services are poised to reinvent the process of navigation and tracking of people and assets in a variety of indoor and subterranean environments. The growing ownership of computationally capable smartphones has laid the foundations of portable fingerprinting-based indoor localization through deep learning. However, as the demand for accurate localization increases, the computational complexity of the associated deep learning models increases as well. We present an approach for reducing the computational requirements of a deep learning-based indoor localization framework while maintaining localization accuracy targets. Our proposed methodology is deployed and validated across multiple smartphones and is shown to deliver up to 42% reduction in prediction latency and 45% reduction in prediction energy as compared to the best-known baseline deep learning-based indoor localization model.

## 1 Introduction

The commercialization of GPS technology in the 1980's completely reformed the transportation industry, simplifying the process of navigation for large ships and airplanes which were dependent on less reliable maps and compasses at the time. A major turning point was the development of the digital GPS ASIC created by Rockwell International, in 1988, using gallium arsenide (GaAs) semiconductor technology [1]. This enabled the first ever handheld GPS receiver to be produced for military applications. Further improvements over the next two decades led to the ubiquitous integration of GPS technology into mobile phones [2]. This empowered individuals to the point where bulky printed maps were no longer needed by automobile drivers and revolutionized outdoor terrestrial navigation around the globe.

Today, increasing capabilities of smart mobile devices are at a tipping point where they can now support localization and navigation technology within indoor environments, which promises to further remold the way humans interact within indoor spaces. As GPS signals cannot penetrate through into indoor locales, highly computationally expensive methods are required that can continuously capture and process wireless signals to support localization. Fortunately, inexpensive and ubiquitously owned smartphones today are computationally capable enough to support high-complexity machine learning models that can be fed by a dense suite of high-fidelity wireless interfaces and sensors on the device. Many researchers are pushing the boundaries on state-of-the-art design optimizations to achieve high-accuracy and real-time indoor localization capabilities on smartphones.

The advances in the indoor localization and navigation domain over the past decade have enabled new commercial and medical applications. Several solutions and standards are being recognized today to enable indoor localization in the public sector. A recent example is the new standard for WiFi that was established in collaboration with Google [3]. The new standard would allow anyone to set up their own localization system by sharing their indoor floor map and the WiFi router positions on that map with Google. Nowadays, large companies such as Amazon and Target have enabled localization services within stores, such that customers can navigate their way to the appropriate products [4]. Several startups have developed their own versions of indoor localization solutions that allow users to employ their smartphones to perform real-time localization. This opens up several interesting use-cases, e.g., for guided tours at museums [39], navigating students and faculty to rooms on a campus building [50], and maneuvering individuals to the closest building exit in case of emergencies [51]. However, such commercial applications require high-quality (high accuracy and low response time) localization solutions and custom hardware components to be deployed at the target location, which drives up the setup and deployment costs.

One way to limit the setup and deployment costs associated with indoor localization services is to use relatively reliable wireless signal sources that are available freely. Due to the boom in the internet and network connectivity across the world in the previous decade, WiFi routers (access points) have become essential and a commonplace feature within indoor locales such as malls, warehouses, hospitals, and schools. Consequently, several recent efforts have focused on delivering high-accuracy localization and navigation solutions for the indoors through a technique called Wi-Fi fingerprinting. Note that fingerprinting is an approach that is applicable for localization in both indoor and outdoor environments, although it is more widely used for indoor environments, whereas trilateration-based approaches (e.g., GPS) are more common for outdoor environments.

Indoor WiFi fingerprinting is based on the idea that each indoor location exhibits a unique signature that is comprised of WiFi signal strengths from visible WiFi routers at that location [5]. Such WiFi Access Point (WAP) Received Signal Strength Indicator (RSSI) values, along with the MAC IDs of these WAPs are captured across various indoor locations during a preprocessing phase, and used to train a model (e.g., machine learning based) that can be deployed on mobile devices. Post-deployment, this model can be used to predict a precise indoor location, given the WiFi RSSI and MAC ID values observed at the location. Alternatively, localization techniques have been proposed that are based on some form of distance relationship between the signal source and destination, such as triangulation [9] and trilateration [10]. However, these approaches suffer from weak wall penetration, multipath fading, and shadowing effects in real-world environments, making it difficult to establish a direct mathematical relationship between RSSI and distance from WAPs. By eliminating this distance relationship between the computed user location and WiFi signal source, WiFi fingerprinting with machine learning models is able to overcome the aforementioned challenges. Fingerprinting also has the advantage of not requiring knowledge of the precise locations of WAPs in an indoor locale, enabling non-intrusive localization.

The overall performance of a machine learning-based indoor localization and navigation framework can be evaluated through metrics such as accuracy, response-time, and scalability of the covered area. Further, the indoor localization framework may be subject to design constraints such as energy consumption, sensor type, and sensor resolution (fidelity). These constraints can be highly stringent when the indoor localization frameworks are deployed on off-the-shell commodity smartphones which have a limited energy budget and utilize severely power-limited processors. While simpler machine learning models such as K-Nearest-Neighbors (KNNs) and Support Vector Machines (SVMs) are scalable, and incur lower energy costs and response times, these models have been shown to be outperformed by more complex and computationally expensive models such as feed-forward deep neural networks (DNNs) and convolutional neural networks (CNNs) that have higher response (inference) times. Moreover, it has been shown that increasing the depth of these neural networks leads to significantly improved localization accuracy but this comes at a cost of higher response times and energy.

Therefore, there is a compelling motivation for designing deep-learning based indoor localization frameworks with a focus on the optimization of their respective deep-learning models such that we can strike a balance between their response-time, energy and achievable localization accuracy. In this paper, we present a novel approach for optimizing Convolutional Neural Networks (CNNs) for indoor localization that can be deployed on mobile devices towards the goal of meeting accuracy requirements (best achievable accuracy through state-of-the-art techniques), while minimizing response times. Our novel contributions in this work are as follows:

[1] We conduct an in-depth experimental analysis on the impact of CNN model depth on an indoor localization framework in terms of the achievable prediction latency and localization accuracy;
[2] For the first time, we adapt and explore the paradigm of conditional computing in the context of deep learning based indoor localization frameworks;
[3] We propose a novel localization framework that can dynamically adapt to the accuracy and latency needs of the target mobile platform at run-time;
[4] We compare the performance of our proposed technique against state-of-the-art deep learning based indoor localization framework over a diverse set of target mobile devices and indoor environments.

## 2 Background and Related Work

### 2.1 Received Signal Strength Indicator (RSSI)

RSSI is a measurement of the power of a received radio signal transmitted by a radio source. The RSSI is captured as the ratio of the received power ($P_r$) to a reference power ($P_{ref}$, usually set to 1mW). The value of RSSI is reported in dBm and is given by:

$$RSSI\ (dBm) = 10 \cdot log \frac{P_r}{P_{ref}} \qquad (1)$$

The received power ($P_r$) is inversely proportional to the square of the distance (*d*) between the signal transmitter and signal receiver in free space and is given by:

$$P_r = P_t \cdot G_t \cdot G_r \left(\frac{\lambda}{4\pi d}\right)^2 \qquad (2)$$

where ($P_r$) is the transmission power, $G_t$ is the gain of transmitter, $G_r$ is the gain of receiver, and λ is the wavelength. This inverse relationship between the received power and distance has often been used by researchers to localize wireless receivers with respect to transmitters at known locations, e.g., estimating the location of a user with a WiFi capable smartphone from a WiFi WAP. However, the free space models based on equations (1) and (2) do not extend well for practical applications. In real environments, the propagation of radio signals suffers from attenuations and interference due to multipath propagation from signal scattering, reflection, and diffraction on obstacles (such as walls, furniture, equipment, people, etc.). Such multipath and shadowing effects cause unpredictable variations in RSSI values at the receiver, thereby severely degrading the performance of free space model based indoor localization approaches, to the point of rendering them impractical for direct use [7].

### 2.2 Indoor Localization Methodologies

Since the inception of wireless radio frequency (RF) based localization a couple of decades ago, a considerable amount of progress has been made in this domain. Here we summarize some of the most noteworthy advancements in this area.

An RF based indoor localization framework, such as one relying on WiFi RF signals, can be classified into three broad sub-domains, i) static propagation based, ii) trilateration or triangulation based, and iii) fingerprinting-based.

*Static propagation* model based techniques are established on the idea that there is a direct correlation between the source signal strength and the distance at which this signal strength is measured. This concept is implemented at design-time by first making signal strength measurements at constant distance intervals from a source in a straight line. The drop in the signal strength in relation to the distance is captured as a static propagation model [8][9][38]. Finally, at run-time the location of the user is predicted based on the received signal strength that will correspond to a specific distance from source value in the model. Such static propagation models are only known to work under extremely controlled conditions with open areas, and no activity. They are often used in conjunction with an error correction system such as Bayesian filters [8] or additional receivers [9] that recalibrate the model over time. However, the RF signal propagation path between every user and source may be unique due to interactions with objects around them. Further, RF transceiver characteristics can vary across devices and manufacturers, which adds to the scalability issues and unpredictability of such models in real-world environments.

*Triangulation and trilateration-based techniques* utilize multiple signal transceivers (e.g., WAPs) to locate people or assets in an indoor environment. They use distance measurements at run-time (by measuring time of flight) such as the distance between multiple WAPs and a mobile device (trilateration) [3][10], or the angles at which the signals from two or more WAPs are received (triangulation) [11]. These techniques have shown to deliver higher accuracy and stability than static propagation models. The techniques can also tolerate device heterogeneity induced uncertainty, to a limited extent (albeit at a high maintenance cost for hardware and software support) [32]. However, these techniques (including Google's RTT [3]) have several limitations, e.g., they need physical locations of all WAPs which is information that may be difficult (or impossible) to obtain in many indoor locales; they require strict clock synchronization among WAPs and the receiver which is not easy to consistently achieve over time; and they may need sophisticated transceivers that are not available in most commodity mobile devices and WiFi Access Points. These techniques also do not work well due to signal interactions with objects in the environment that induce signal multipath, shadowing, and variation in propagation speed through materials other than air [12].

*Fingerprinting techniques* and their viable implementations can utilize machine learning domain to overcome the aforementioned challenges associated with signal interactions and maintenance costs. Our work therefore utilizes this approach. We discuss prior work in this area in the next sub-section.

## 2.3 Fingerprinting-based Indoor Localization

Due to the limitations of the static propagation model-based and triangulation or trilateration-based techniques, researchers are now increasingly focusing on fingerprinting-based indoor localization techniques. Fingerprinting can be implemented in two ways: 1) custom infrastructure based: where custom AP beacons are installed in indoor environments based on Ultra-Wide Band (UWB) [12], Bluetooth [13] or Zigbee [14], and 2) infrastructure-free: where freely available signal sources such as earth's magnetic field [15] and WiFi [16] are utilized. The former approach lacks scalability and suffers from high costs. Moreover, smartphones do not have transceivers for protocols such as UWB and Zigbee. The latter approach, because of its low cost and ease of setup, is therefore more preferable.

A generalized view of an infrastructure-free WiFi RSSI fingerprinting-based indoor localization framework is presented in figure 1. Such frameworks usually consist of two phases: the offline (or training) phase and the online (or testing) phase. From figure 1, we note that the offline phase consists of the user collecting WiFi fingerprints to create an RSSI fingerprint database. Each row of this database consists of RSSIs for various WAPs observed at a given location (reference point). The row of RSSI information is also known as an RSSI vector. One may collect RSSI vectors at each reference location multiple times (e.g., at different times of the day or week) to capture a broader range of RF signal behavior at that location. It is important to note that while fingerprinting-based indoor localization requires cumbersome collection of fingerprints over a large area (RSSI database), a significant body of work has been dedicated to addressing this challenge, e.g., [41-44]. Once the RSSI database is constructed, it is then used to train a Machine Learning (ML) model, where the RSSI vector is the input and the reference location is the output of the model. This model is finally deployed on to the mobile device that will be used by the end user for indoor localization. As discussed before, deploying indoor localization ML models on smartphones is becoming a common practice due to various infrastructure costs and computational capability benefits.

In the online or testing phase, as shown in figure 1, the target mobile device captures an RSSI vector as a user moves across an indoor space. The RSSI vector is then fed to the ML model on the mobile device, that in turn predicts the location of the user. This process of capturing RSSI vectors and then predicting the user's location continues to occur in a cyclic fashion to create an ongoing stream of location prediction cycles. The time taken to complete a prediction cycle (i.e., prediction latency) and the accuracy of the predicted location are two key metrics that describe the responsiveness and effectiveness of an indoor localization framework. A truly real-time localization framework is expected to be responsive to the user's movement, providing continuous predictions (approximately taking no more than a few tens of milliseconds for each prediction), while maintaining an acceptable level of location prediction accuracy.

Over the previous decade, the area of fingerprinting-based indoor localization has been heavily explored. UjindoorLoc [16] describes a technique to create a WiFi fingerprint database and employs a KNN (K-Nearest Neighbor) based model to predict location. Their average accuracy using KNN is 7.9 meters. Radar [17] and Indoor Atlas [18] are early works that proposed using hybrid indoor localization techniques. Radar [17] combined inertial sensors (dead reckoning) with WiFi signal propagation models, whereas Indoor Atlas [18] combined information from several sensors such as magnetic, inertial, and camera sensors, for indoor localization. LearnLoc [19] combined non-deep ML models with inertial sensor data and WiFi fingerprinting to propose a framework that trades-off indoor localization accuracy and energy efficiency on smartphones.

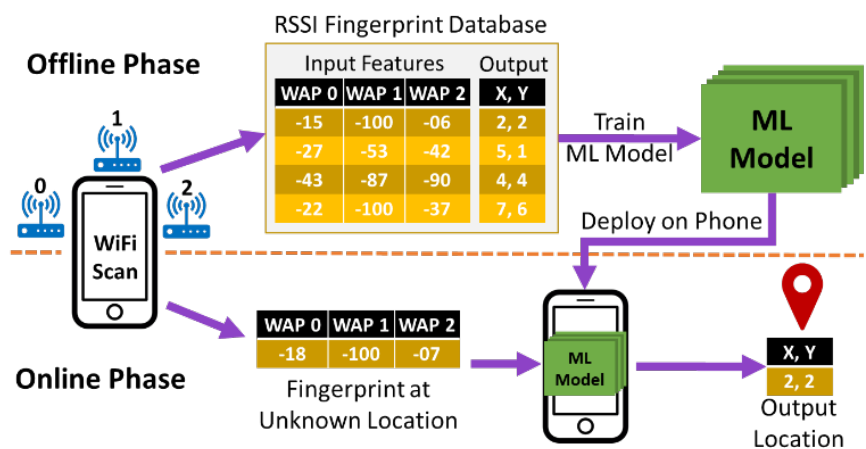

**Figure 1: A generalized overview of the online and offline phases of fingerprinting-based localization frameworks**

As the computational capabilities of smartphones have increased in recent years, researchers have begun to explore the possibilities of deploying more complex algorithms such as DNNs on mobile devices towards the goal of attaining higher localization accuracies. Publicly available neural network frameworks such as TensorFlow and PyTorch have enabled rapid prototyping of complicated deep learning models and can be deployed on mobile devices with ease. The work in [20] presents an approach that uses DNNs and Hidden Markov Models (HMMs) for WiFi RSSI fingerprinting. DeepFi [21] and ConFi [22] propose approaches that use the Channel State Information (CSI) of WiFi signals to create fingerprints. But the CSI information in these approaches was obtained through the use of specialized hardware attached to a laptop. None of the smartphones available today have the ability to capture CSI data. Due to this limitation, it is not feasible to implement these techniques on smartphones. Deep Belief Networks (DBN) [23] have also been used for indoor localization, but the proposed technology is heavily reliant on custom UWB beacons that lead to a very high implementation cost. The work in [40] presents a deep-learning-based indoor localization framework that fuses fingerprints from two sources: WiFi, and magnetic signals, to produce the user's location estimate. *A limitation of all of these prior works on deep learning and fingerprinting based indoor* *localization is that they focus solely on indoor localization accuracy, without considering the responsiveness (i.e., prediction latency performance) of the proposed frameworks.* The responsiveness of a fingerprinting-based indoor localization framework is heavily dependent on the prediction latencies of their respective machine learning models, as well as the specific mobile device platform that the model is deployed on. True real-time indoor localization can only be achieved if the time to sample signal fingerprints and producing a location prediction is small enough that there is no lag between user movement and location prediction displayed on the user's mobile device.

In summary, existing indoor localization frameworks focus extensively on prediction accuracy, however, very limited attention is placed on architectural optimization of existing deep learning models for lower prediction time and energy. To the best of our knowledge, there are no works in the area of fingerprinting-based indoor localization that explicitly focus on the optimization of deep learning models with an emphasis on reducing response-time with no loss (or gain) in accuracy. These factors are critical to achieve consistent performance and scaling of deep learning based indoor localization frameworks across a variety of mobile devices. Towards the end goal of creating responsive real-time indoor localization frameworks we propose the QuickLoc framework that adapts the early exit deep learning based architectural design philosophies presented in [24] and [25] to the domain of indoor localization, for the first time. QuickLoc has the capability to strike a balance between response time while maintaining high indoor localization accuracy.

## 2.4 Model Compression for Energy-Efficient Deployment

Convolutional Neural Networks (CNNs) are often characterized as being computationally expensive and memory intensive. This is a major limitation when deploying such complex models on resource constrained embedded systems such as smartphones. Such a challenge prompted researchers to develop techniques that either compress or accelerate CNN execution without significant loss in performance. Some well-known approaches for model compression techniques that enables the energy-efficient design of CNNs include quantization, pruning, and conditional computing. Note that none of these compression techniques have yet been explored for the problem domain of indoor localization.

Quantization reduces the number of bits required to represent each weight value within layers of a neural network model. This approach has shown to lower the memory and computational requirements by several orders of magnitude with minimal loss in accuracy, enabling higher energy-efficiency overall [52]. Another popular technique for reducing model size is pruning which involves the removal of weights (biases and activations can also be pruned) based on some pruning criteria [53]. The pruned elements are "trimmed" from the model and do not take part in the computation. Other commonly known compression approaches include knowledge distillation and weight sharing. More details on these can be found in [54] and [55] respectively. Conditional computation is another compression technique that focuses on a neural network design philosophy where different portions of the neural network are activated based on the input fed to the model. The portions of the neural network to be executed is based on some predetermined logic or learned gated structure [24]. The major advantage of activating fewer neural network units is that propagating information through the conditional network would be faster. The concept of conditional computing is used in early exit neural networks [25]. Early exit neural networks are based on the idea that not all input samples require the evaluation of the full model. If satisfactory results are achieved, the computation can be halted before the normal endpoint of the model is reached. A major advantage of utilizing early exit style models is that once the full model has been trained (including early exits) the criteria for satisfactory results (see section 6.3) can be tweaked or tuned. Also, unlike the compression techniques discussed above, no additional computationally expensive model training is required. This allows for greater flexibility and adaptability of the neural network model post-deployment. More details on conditional computation and early exits are presented in section 5.

## 3 CNNLOC Framework Overview

In this section, we discuss the concepts associated with the WiFi fingerprinting-based indoor localization framework proposed in [26], called CNNLOC. We utilize CNNLOC as the baseline work due to the benefits of using WiFi RSSI only indoor localization and deep learning as highlighted in the previous section, as well as due to the fact that this recent work has been deployed on mobile devices and shown to outperform other solutions in the indoor localization problem domain. Our goal in this work is to improve upon the performance achievable by CNNLOC. However, please note that the design methodology proposed in this paper can be applied to other deep learning frameworks as well, such as [20]-[23][40].

## 3.1 Convolutional Neural Networks

Convolutional neural networks (CNNs) are a form of deep neural networks that are specially designed and optimized for image classification. They have been shown to deliver significantly higher classification accuracy as compared to conventional fast-forward only DNNs due to their enhanced pattern recognition and feature extraction capabilities.

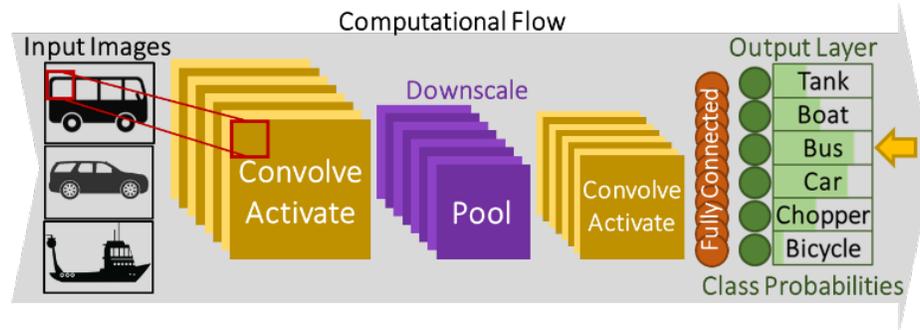

**Figure 2: An example of a Convolutional Neural Network (CNN) design**

As shown in figure 2, a typical CNN model has three main functional components (or types of layers): convolutional layers (that perform "convolve" operations), pooling layers (that "pool" or downsample the activations), and fully connected layers (that feed flattened data to the output for predictions). Convolutional and fully connected layers also have associated activation functions ("activate" operations) whose role is to introduce non-linearity into the neural network model, allowing the model to learn complex, non-linear patterns. ReLu (Rectified Linear Units) and its variants (such as leaky ReLu) are the most popular activation functions in the pattern recognition domain.

In general, CNN models learn patterns in images by focusing on small sections of the image, known as a frame, as shown in figure 2. The frame moves over a given image in small strides. Each convolutional layer consists of filters (matrices) that hold weight values. The layer involves convolutional operations (dot products) performed between the current input frame and filter weights followed by passing the result through the activation function. The pooling layer is responsible for down sampling the output from a convolutional layer, thereby reducing the computational requirements by the next set of convolution layers. A set of fully connected layers is typically utilized after potentially multiple convolutional and pooling layers, to reduce the depth of activations (i.e., data propagating through the CNN) before a final classification can be performed. Typically, a SoftMax activation function is applied to the output of the last fully connected layer, to generate the probability distribution for the classes being predicted by the model. In the testing (or inference) phase of a CNN model, an image is fed to the model which in turn produces class probabilities. The class with the highest probability is identified as the output prediction. Further details on the design of CNNs can be found in [27] and [28].

## 3.2 Indoor Localization with CNNLOC

The CNNLOC indoor localization framework [26] consists of two major components in the offline phase. The first component involves capturing the RSSI fingerprints for different indoor locations, and then converting each RSSI fingerprint vector that is associated with a location (reference point) into an image associated with the same location. The second component of the offline phase is the training of a CNN model using the images created from RSSI vectors. In the online phase, the same process is used to create an image (based on observed RSSI values), which is fed into the trained CNN model for location prediction.

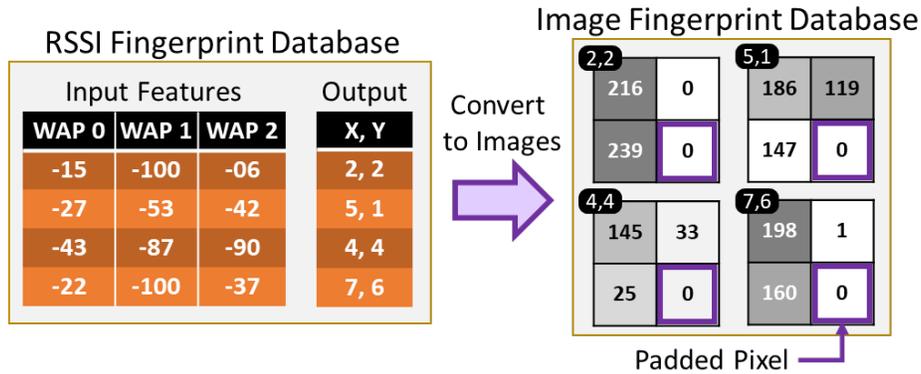

Figure 3: Converting RSSI fingerprint vectors to RSSI images

A simplified overview of the process of converting an RSSI fingerprint vector into an image is shown in Figure 3. The RSSI vector consists of RSSI values in the range of -100 to 0 dBm (low signal strength to high signal strength). These values are normalized to a range of 0 to 255, which corresponds to the pixel intensity on the image. The dimensions of the RSSI image are set to be the closest square to the number of visible WAPs on the path. For example, in figure 3, the RSSI vector has a size of 3, and the closest square would have 4 pixels in it, therefore, the dimensions of the image are set to 2×2. A pixel with zero intensity is padded at the end to increase the size of the vector as shown in figure 3. The generated image then becomes a part of the offline database of images used to train the CNN model.

In the online phase, this same process of image creation is used with the RSSI vector observed by the user at any location, and the resulting image is fed to the trained CNN model to get a location prediction. It is important to note that in the online phase of CNNLOC, the input image will always remain the same size as in the offline phase, such that each pixel in the image corresponds to the RSSI value from a WAP with a specific MAC IDs. In case a specific MAC ID observed in the offline phase is no longer visible in the online phase, we set the RSSI value for it to -100 dBm. This results in the pixel value corresponding to that MAC ID being set to zero.

## 4 Localization Inference Analysis

We begin with an analysis of the impact of model depth on the state-of-the-art indoor localization framework CNNLOC [26] described in the previous section. To capture the impact, we train three unique CNN models for the paths shown in figure 4 and deploy them on the four mobile devices summarized in table 1. The first model has only one layer of convolution, the second model has two layers, and the third model has three layers of convolution. Due to small input image sizes, our models do not have pooling layers [26]. It is important to note that each of these models are trained to cover all of the paths shown in figure 4. More details about the indoor paths and devices are covered in Section 7.2. Further, to curtail the complexity of this experiment, we utilize the same hyperparameters for the convolutional layers as in the model described in section 6.

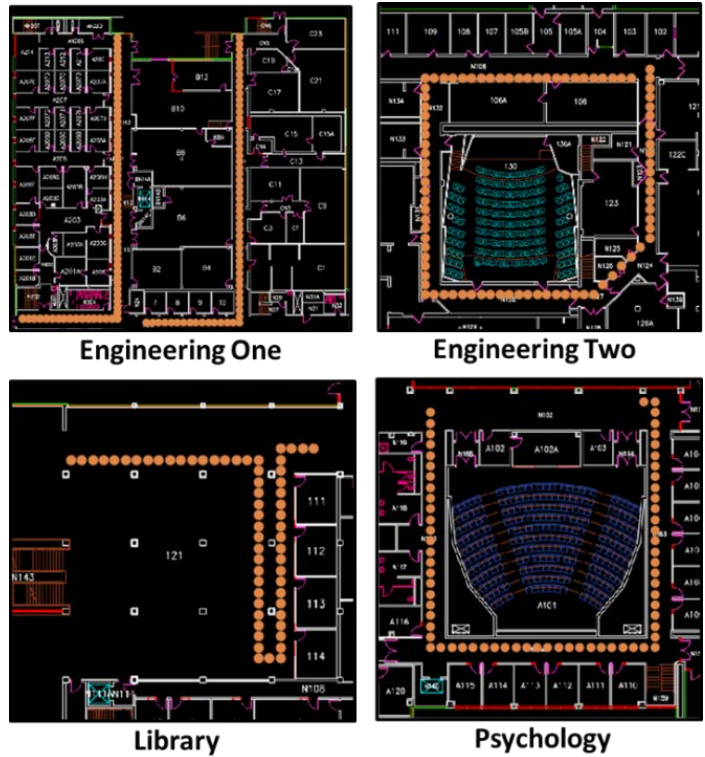

**Figure 4: Indoor paths in different buildings for indoor localization analysis. Reference locations (where RSSI values were recorded to train the CNN models) along the indoor paths are indicated by orange dots**

**Table 1: Details of smartphones used in experiments**

| Smartphone | Chipset | CPU Freq. | RAM | Battery Capacity (mAh) |
|---|---|---|---|---|
| OnePlus 3 (OP3) | Snapdragon 820 | 2350 MHz | 6 GB | 3000 |
| Moto Z2 (MZ2) | Snapdragon 835 | 2350 MHz | 4 GB | 2730 |
| Samsung S6 (GS6) | Exynos 7420 | 2100 MHz | 3 GB | 2550 |
| Samsung S7 (GS7) | Snapdragon 820 | 2300 MHz | 4 GB | 3000 |

Figure 5 depicts the variation of model prediction accuracy and average latency for CNN models of varying depths deployed on the four different mobile devices. Considering the fact that smartphone chipsets are usually heterogenous in nature and consists of complex cores (clocked at higher frequencies) and simpler cores (clocked at lower frequencies), we report the latency values for situations where the model is specifically executed on the core clocked at the highest frequency available. For each CNN model depth increment, we added an additional convolutional layer to the model. The most obvious observation is that in general the deepest model incurs significantly higher prediction (inference) latency. The model with three convolutional layers is up to 8x slower (OP3 device) than its shallow single convolutional layer counterpart. By increasing the model depth, we are able to boost the localization accuracy from 85% to 95%. However, this boost in localization accuracy comes at a hefty price of higher localization time. On the other hand, this observation also indicates that the patterns associated with 85% of the fingerprints are easily identifiable and utilizing deeper models is actually inefficient for most of the path covered by the user.

Prediction (inference) latency is a critical factor for the fulfillment of real-time indoor localization and navigation through fingerprinting. This is especially true for hybrid indoor localization frameworks that combine various techniques such as fingerprinting, dead reckoning, and particle filters at the same time to produce consistent high-fidelity results. For example, an indoor localization framework that depends on a machine learning model to inform other subsystems and aims to update the smartphone display every time the user moves by a 10th of a meter, requires the predicted location to update every 30 milliseconds (assuming an average movement speed of 3m/s [29]). This is only achievable if the indoor localization framework is able to pre-process the fingerprint and produce an inference from the deep learning model at a latency that does not exceed approximately 30 milliseconds, based on our empirical experience with deploying and running such models for indoor localization on smartphones. In figure 5, we observe that the 3-layer model is unable to deliver such latency on most mobile devices. We also tested CNN models with more than 3 layers (results omitted for brevity) and found a much higher inference latency with those deeper models, which made them not very well suited to our real-time indoor localization inference time goals.

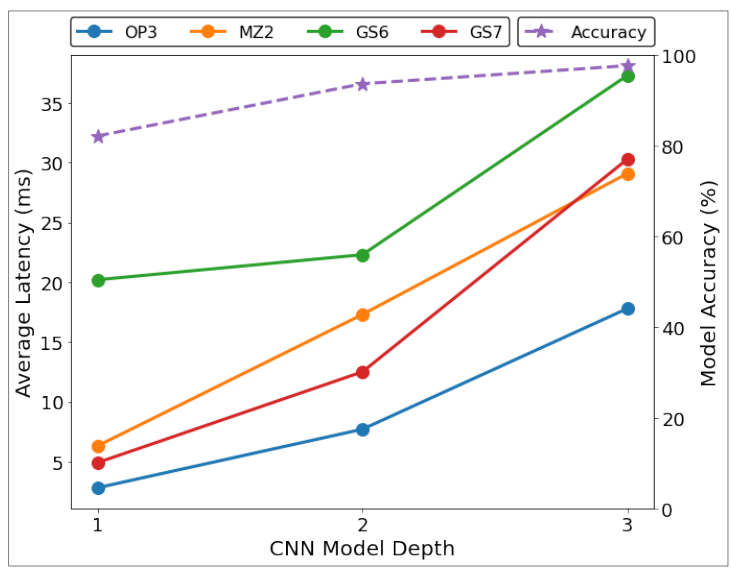

**Figure 5. Relationship between CNN model depth, average prediction latency, and accuracy for the four smartphones shown in table 1**

Another observation from figure 5 is the variation in prediction latency across the various mobile devices. While the latencies of OP3, MZ2 and GS7 devices are similar for a model depth with depth 1, the latencies for the models with a depth of 2 and 3 vary greatly. These localization latencies of the same CNN model are significantly dependent on the specifications and optimizations for the target mobile device. By comparing the device configurations in table 1 and figure 5, we conjecture that the DRAM specifications may play a crucial role in determining the prediction latency of the CNNLOC model. Further, we noted from our analysis that depending on the type of core the model workload is allocated to, the localization latency could be up to 3× worse than the ones reported in figure 5.

As smartphones are powered by batteries and thus, limited by an energy budget, utilizing deep models for a task that can be accomplished using a shallower model wastes computational resources that could have been allotted to other tasks to further improve localization accuracy, such as direction estimation, via sensor fusion with inertial sensors, and using particle (or Kalman) filters. Further, as we scale up the number of reference points on which RSSI readings are measured during the training phase, and the number of WAPs, the model depth and complexity needed to achieve accurate indoor localization will also increase. This will in turn result in higher prediction latencies, which will create a challenge for the deployment of such models on mobile devices.

The observations from the analysis in this section suggest a critical need for indoor localization frameworks that can deliver high localization accuracy without trading off localization latency and that can also perform consistently across a wide variety of heterogeneous mobile devices

## 5 Conditional Early Exit Models

From our analysis in the previous section, we observe that a shallower model is able to predict 85% of the locations accurately. This observation suggests that we do not need to use a deeper model to predict the user's location in every prediction cycle. Even though the deeper model can predict the user's location more accurately on average, it comes at the considerable cost of higher inference latency. Further, as the technique in CNNLOC [26] and other similar techniques are scaled up, the high complexity of the deployed model may become a barrier from its ubiquitous use in resource constrained devices such as smartphones and smartwatches.

Towards the goal of optimizing the inference latency of the indoor localization model, we exploit the observation that a large portion the WiFi fingerprints in the training dataset can be learned easily and effectively by simpler models. However, we also want to ensure that locations that can benefit from a deeper model can actually leverage the benefits of additional layers for improved prediction accuracy. To realize such an implementation, we build on the idea of early exit in deep learning models, as proposed in [24] and [25]. We explore the possibility of branching the computation after each convolutional operation to achieve an acceptable response based on uncertainty sampling methods such as confidence difference, confidence ratio, or entropy. These are discussed later in this section.

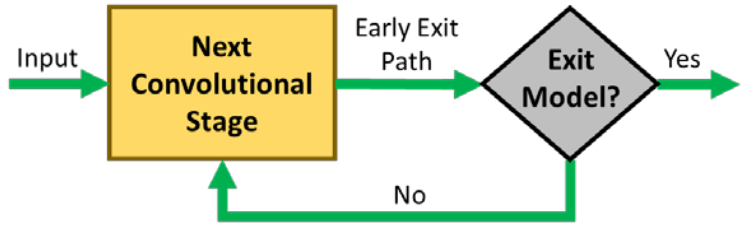

Figure 6: Early exit strategy depicted as a state machine

The adapted conditional exit strategy can be captured as a state machine and is depicted in figure 6. The input to the state machine is an image that is fed to the convolutional neural network. After each convolutional layer or stage, the output is fed to an exit path. The output class probabilities produced at the end of the exit path are then fed to an uncertainty sampling method. The satisfactory result of the uncertainty sampling method is used to recognize the validity of the predicted class at the current exit stage. In case we are confident of the model output at the current exit stage, the current prediction is accepted. In the case that we are not confident of our early prediction, we continue on to the next convolutional stage and evaluate the output of that stage for conditional exit. In this manner, we expect a reduction in the inference time for a majority of the location prediction cycles through uncertainty sampling based early exits.

Consider the example that was shown in figure 1 earlier. The model is fed an input image of a car and produces the probabilities for various vehicle classes, such as a tank, boat, bus etc. While the class probability of a bus is the highest, the probability of the input image being a car is only slightly lower. Such behavior is expected as the images of buses and cars may have similar features such as wheels and large windows. However, a CNN model is likely to easily differentiate between a boat and a car due to dissimilar features or patterns in the images. In this manner, if input images of a CNN model have significantly varying features, they can be easier to identify using shallow models. Further, the distribution of probabilities across the various output classes (as seen in figure 1) can be utilized to capture the model's confidence in its prediction. The class of techniques used for this purpose are known as uncertainty sampling methods. A subset of these methods is explored in this work for our problem of fingerprinting-based indoor localization. The explored methods are described below:

[5] **Least Confidence:** This is the difference between the most confident prediction and 100% confidence;

[6] **Margin of Confidence:** This is the difference between the most confident and the second most confident prediction;
[7] **Ratio of Confidence:** This is the ratio between the top two highest class probabilities (most confident);
[8] **Entropy:** This is a concept derived from information theory that describes the level of uncertainty associated with one possible outcome, compared to all other outcomes [30].

The early exit strategy reduces inference latency by limiting the overall computation required for each prediction (inference). Other well-known techniques such as model compression and quantization are orthogonal to this method and can be applied in conjunction with this approach. Based on the proposed early exit strategy, several predictions follow a shorter path to completion thereby establishing shallower computational paths, with lower latencies. This is also a highly beneficial behavior as shallower models are less likely to be a victim of the vanishing gradient problem [31]. Shallower models in our problem domain are sometimes able to identify some locations accurately that might be harder for deeper models to predict accurately, due to this issue. In our experiments, we found evidence of this phenomenon, where utilizing early exit models allowed for improving localization accuracy under some conditions.

It is important to note that while figure 6 depicts the early exit model behavior as a state machine, it does not capture the specific conditional early exit model design presented in this work. The early exit path depicted in figure 6 may contain one or more neural network layers that are not a part of the original CNN model. We present the detailed process for creating the early exit model that we used in the proposed QuickLoc fingerprinting-based indoor localization framework in the next section.

# 6 QuickLoc Framework

In this section, we discuss the design of our QuickLoc framework for the purpose of reducing inference latency.

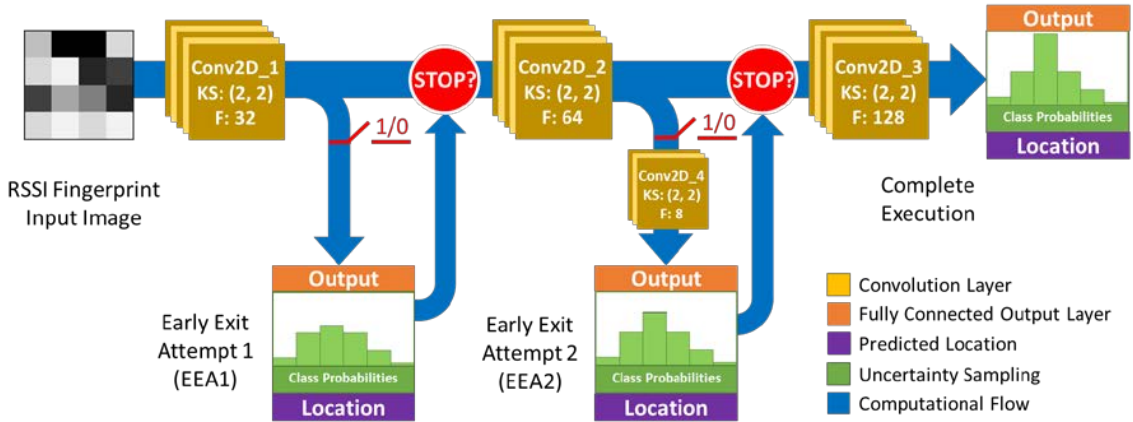

**Figure 7: Overall flow of computation with conditional early exits for the proposed QuickLoc indoor localization framework**

## 6.1 QuickLoc CNN Model Design

The proposed model design for this work is depicted in figure 7 and the number of parameters in each layer is presented in table 2. The baseline model consists of three convolutional layers each with a small kernel size of 2×2 and a stride size of 1. A small kernel size is chosen as the RSSI fingerprint images have a small resolution as discussed in CNNLOC [26]. We further utilize the same filter size in each layer to maintain simplicity in this exploration. A real-world deployment could have different filter sizes at each convolutional layer. The baseline model is designed such that the number of filters is increased as the depth of the model increases. This forces the CNN model to learn increasing number of complex features as the model depth increases. The first convolutional layer "Conv2D_1" consists of 32 filters producing only 160 parameters, followed by the second layer "Conv2D_2" with 64 filters (8.2K parameters) and finally, the third convolutional layer "Conv2D_3" consists of 128 filters (32.8K parameters). Each convolutional layer is followed by a ReLu activation function, as in [26].

Based on our discussion in the previous section, we attempt to perform an early exit after each convolutional stage. The first early exit attempt (EEA1) comes after Conv2D_1 and only consists of a single output layer. Each fully connected output layer consists of 342 neurons (same as the total number of reference points) followed by the Softmax activation function. In EEA2, an additional convolutional layer "Conv2D_4" with only a few filters (8 filters producing 2K parameters) is attached before the output layer. From table 2, we observe that all of the output layers consist of a large number of parameters. As the QuickLoc model adds multiple output layers to the baseline model, the resulting model is expected to have a larger memory footprint. For a more realistic deployment, in the early design phases of the model, the designer might start with N-1 EEAs. However, only some of them might add value to the overall application design. For example, in a model with 10 total CNN layers, the first 6-7 early exits may yield very low localization accuracy. Such low accuracy EEAs may not add to the accuracy attained through supplementary techniques (particle filters, dead reckoning). The number and choices of EEAs to keep are up to the application designer. These choices will also have a critical impact on the model footprint. An analysis into memory footprint at run-time is presented in section 8.5. Further, the hyperparameter selection of the two early exit branches is discussed in the next subsection.

**Table 2: Number of parameters in the QuickLoc model**

| QuickLoc Layer | Number of Parameters |
|---|---|
| Conv2d_1 | 160 |
| EEA1 Output | 9,204,246 |
| Conv2d_2 | 8,256 |
| Conv2d_4 | 2,056 |
| EEA2 Output | 1,994,886 |
| Conv2d_3 | 32,896 |
| Output | 31,913,046 |

## 6.2 QuickLoc CNN Model Training

The training process for the model presented in figure 7 begins with the baseline CNN models design and training as discussed in [26]. To highlight the full potential of our proposed technique we chose to train a single model for all of the buildings in our dataset instead of a model for each building.

Once the baseline model is established, the first early exit stage (Conv2D_1+ EEA1) is created by training the layers on the EEA1 path such that the weights associated with the convolutional layers of the baseline model (Conv2D_1) are frozen and remain unchanged in the training process. Once the layers associated with the exit path have been trained, they are manually attached to the full baseline model. This process is repeated for each convolutional layer in the baseline CNN model.

While designing the layers on each early exit path, two design philosophies are followed. The first is that the depth of the early exit itself is generally directly proportional to the depth of convolutional stage whose output is fed to the early exit. In this manner, we note that EEA2 is computationally more expensive than EEA1. The second is that the computational expense of an early exit should be significantly lower than the remaining computation in the baseline model. It is important to note that the expense of a computational path is dependent on several factors such as number of layers, number of parameters in each layer, and the types of layers. The proposed design in this work considers all of these factors.

## 6.3 Uncertainty Sampling Threshold

At each early exit attempt, the confidence associated with the predicted output is calculated through class probabilities using one of the various uncertainty sampling techniques presented in the previous section. If the uncertainty of the predicted class is within an acceptable threshold, the location prediction at the current early exit is accepted. The value of these thresholds and the acceptable range is dependent on the type of uncertainty sampling method used. A sensitivity analysis on the choice of the uncertainty sampling technique is presented in the experimental section (section 8).

## 6.4 Post-Deployment Configuration Adaptivity

From our analysis shown in figure 5, we observe that the performance of a CNN model can vary significantly across different devices. Subtle variations such as SoC type and DRAM size can lead to significant performance variations for the same CNN model. Due to this behavior, a one-for-all CNN model solution is inefficient and is likely to deliver inconsistent inference time on new devices not evaluated in the training phase.

Another notable challenge of the proposed early exit strategy is the computational or latency penalty due to inferences that are unable to confidently exit on any of the early exit attempts. The latencies associated with inferences or location predictions that completely fail to exit early would be generally greater than the baseline CNN model without early exit attempts.

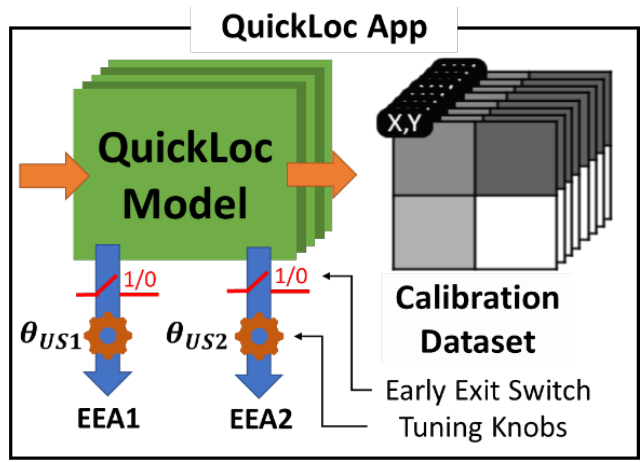

Figure 8: Contents of QuickLoc app package depicting tunable uncertainty sampling threshold ($\theta_{US}$) and early exit switches as configurable parameters.

To overcome these challenges, we implemented the capability of enabling or disabling early exit paths once the model has been deployed on a smartphone and is in the testing phase. This is due to the fact that there may be multiple combinations of the ways the proposed early exit CNN model in the QuickLoc framework can be configured. For example, a model that only has EEA1 enabled, may deliver higher accuracy and lower inference time than a model with both EEA1 and EEA2 enabled. Once the model has been deployed on a smartphone, it undergoes a self-configuration process using a limited set of RSSI fingerprints and associated reference points to identify an early exit configuration and uncertainty sampling threshold that delivers the best results.

Figure 8 depicts the various components of the QuickLoc indoor localization application (testing phase) and associated tunable parameters. The application also consists of a small set of labeled training data used for calibrating the various control knobs of the QuickLoc model. For each EEA, there are two control parameters: self-enable/disable switch and uncertainty sampling threshold value. Once the application is installed on a smartphone in the testing phase, the labeled training data is utilized to identify a localization error and inference latency for each possible configuration of the QuickLoc model (EEA and $\theta_{US}$). This would allow us to identify multiple configurations that deliver higher accuracies than the baseline (no early exit) at lower inference latencies. We present an analysis later in section 8.2 that uses this approach to explore multiple configurations of the model across different smartphone devices.

For the purpose of this work, the default configuration is the one that produces a reduction in prediction latency with no loss in localization accuracy. However, in practice, the QuickLoc configuration can also be adjusted on-demand to meet specific latency goals at run-time. The benefit of such an approach is the ability to trade off latency with accuracy in the testing phase. For example, when using QuickLoc in combination with dead reckoning, one may choose to change the QuickLoc configuration with higher inference latency and lower localization error at run-time if the user is detected to be moving slower. To understand the impact of the various early exit configurations we present a sensitivity analysis on various devices later in section 8.

# 7 Experimental Setup

## 7.1 Heterogenous Device Specifications

To capture the variation in performance across heterogeneous devices, we first design and train the QuickLoc model based only on the OP3 device characteristics and then deploy our indoor localization model onto three other smartphones with unique hardware specifications in the testing phase. The specifications for each of these devices is captured in Table 1. This allows us to explore the impact of device specification heterogeneity such as DRAM and SoC type that can impact localization latency. Such a model design and training process is adopted to simulate a real-world scenario where the specifications of the target mobile platform may be unknown when deploying QuickLoc on new device.

It is important to note that we do not consider the impact of RF-based heterogeneity (due to different radio antennas) across smartphones. The purpose of this analysis is to evaluate the improvements in energy-efficiency and latency achieved through QuickLoc. There have been previous works in the domain of fingerprinting-based indoor localization that solely address the challenges associated with RF-based heterogeneity such as [45-49]. These works either utilize metrics that are resilient to device heterogeneity or standardize the fingerprints before they are fed to a model. Such concepts can be extended and applied together with QuickLoc.

## 7.2 Indoor Paths for Localization Benchmarking

We compare the localization accuracy and latency for the proposed QuickLoc framework using a benchmark dataset with 342 reference locations. The benchmark spans over a large university campus with varying environmental conditions and WiFi WAP densities. The dataset covers four buildings. The paths within these buildings are shown in figure 4; with each orange dot indicating a reference point on a path within the building that is one meter apart. The paths vary from 70 to 90 meters in length and the number of visible WAPs along these paths varies between 78 to 218. The process of capturing fingerprints employed for the purpose of training and testing within a building is completed within a couple of hours. We returned to collect data in these buildings at different times and performed post-processing on the collected data to eliminate temporary WAPs, e.g., mobile hotspots created by individuals in the buildings.

The path sections in Engineering Building One consist of labs, mechanical equipment, and office spaces. This path was specifically chosen as it has the largest amount of electrical and magnetic devices in its vicinity, that interacts with WiFi signals to produce noisy fingerprints. The psychology and library buildings were recently renovated with a mix of wooden and metallic structures in its surrounding environment. The path sections are mostly surrounded by large halls and classrooms such that the impact of multi-path effects and shadowing is relatively lower as compared to other buildings. Finally, the last building covered is an engineering building (Engineering Two). This is the most versatile path section covered. It is one of the oldest buildings on campus and mostly constructed of wood and concrete. The building consists of labs with metallic equipment, office spaces, and large classroom halls. The reference points for fingerprints over all buildings are 1-meter apart. Ten fingerprint samples per reference location were collected. The WiFi fingerprints in this benchmark were captured in the offline phase while holding the smartphones at an average height of 1.5 meters above ground such that the device screen is zenith facing. Testing (online phase) was performed by 5 users with heights varying between 175-192 cm. The users held the device close to their chest height while facing the smartphone display.

## 7.3 Comparison with Previous Work

The performance of QuickLoc is compared to its non-early exit capable counterpart CNNLOC [26], which is the baseline model in our analysis. Additionally, we compare QuickLoc with conventional machine learning indoor localization frameworks that utilize K-Nearest Neighbor (KNN) [19] and Support Vector Regression (SVR) [35]. The KNN-based indoor localization framework [19] algorithm is based on the idea that the RSSI fingerprints at a given reference point would be close to each other in the Euclidian space. The SVR-based framework [35] attempts to create a set of hyperplanes, based on groups of RSSI fingerprints, each associated with a specific reference point. These frameworks utilize relatively light-weight machine learning algorithms that lead to lower inference latency. The purpose of comparing QuickLoc against the works in KNN [19] and SVR [35] is to contrast the inference latency and accuracy of QuickLoc against known light-weight indoor localization platforms.

## 7.4 Deployment and Evaluation

The early exit model is trained as described in section 6.2. The uncertainty sampling methods and threshold values for each early exit was empirically evaluated and set based on the OP3 device in the offline phase. The trained early exit model and the baseline models are deployed on smartphones using an Android app with timers for capturing latency. Once deployed, QuickLoc automatically reconfigures itself for the target smartphone. This is a one-time process that occurs at the first launch of the QuickLoc app.

We deployed the QuickLoc on smartphones using Tensorflow Lite and used the official C-based benchmarking application [36] over the Android Debug Bridge (ADB) to capture latency and memory requirements. This allows us to minimize the impact variations produced by the Android OS application manager layer. The energy analysis presented in section 8, is conducted by capturing battery drain characteristics attained using the BatteryManager API for Android [37]. We do not perform any form of post-training quantization on our Tensorflow Lite models. However, doing so would only further improve the inference latency of the QuickLoc model at the cost of localization accuracy. Lastly, WiFi RSSI fingerprint scans took anywhere from 1.5 to 4 seconds, depending on the smartphone being tested. As we move towards the eventual goal of real-time localization, higher sampling (scan) rates are needed. Recent efforts to enable monitor mode for WiFi chipsets for smartphones are a step in that direction, by enabling more frequent packet-by-packet updates to WAP RSSIs [33][34].

# 8 Experimental Results

## 8.1 Sensitivity Analysis for Uncertainty Sampling

In this subsection, we present results for a sensitivity analysis on the type of uncertainty sampling technique and its associated threshold value for our proposed QuickLoc model. The sensitivity analysis is conducted on the OP3 mobile device as it shows the least variation in prediction latency (figure 5). Through the selection of this device we intend to describe the performance of QuickLoc on a smartphone whose prediction latency is the least flexible and hence, is expected to produce the least improvement. For simplicity, we utilize the same threshold values for EEA1 and EEA2 (both enabled).

Figure 9 presents the average localization errors and the prediction latencies on the left and right vertical axes, respectively; and the uncertainty threshold values on the horizontal axes, for the four uncertainty sampling techniques described in section 5 (margin of confidence, least confidence, ratio of confidence, and entropy). The dashed horizontal red lines and green lines represent the localization error and latencies (respectively) for the baseline CNNLOC framework. We observe that the performance of QuickLoc is greatly impacted by the choice of uncertainty sampling method. In figure 9, we observe that the Least Confidence method performs the worst as there are no configurations of the uncertainty threshold value for which QuickLoc delivers higher accuracy at a lower latency than the baseline CNNLOC model. In contrast, Margin of Confidence and Entropy produce the most configurations with both improved latency and localization accuracy. Due to the logarithmic nature of localization error for the entropy method, it may not be the best choice for a framework variant that throttles the uncertainty threshold for a tradeoff between localization accuracy and latency. More analysis on this subject is presented in a later subsection. From the analysis presented in this section, the margin of confidence is the best choice for the OP3 device. However, the appropriate adaptive configuration for each device may be unique for QuickLoc in the online phase on the target smartphone. The next sub-section highlights QuickLoc's configuration flexibility for various devices.

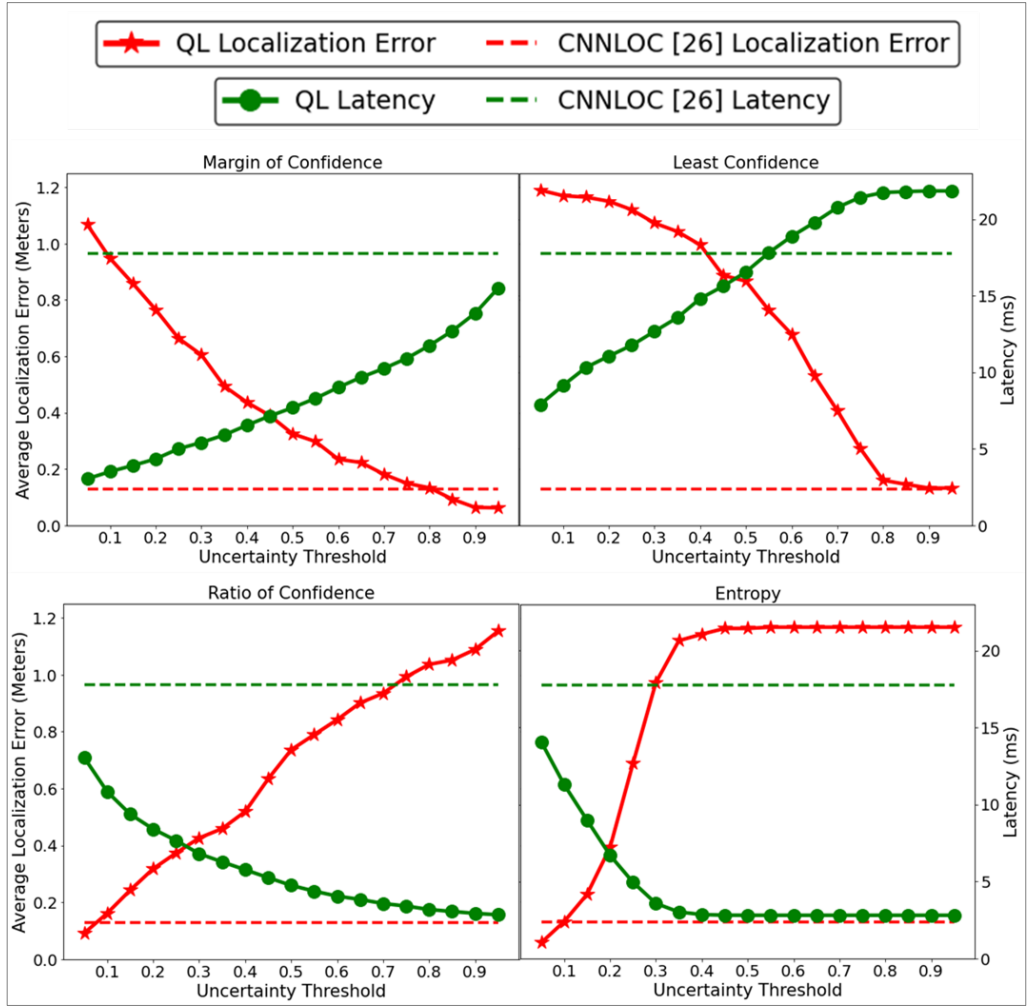

**Figure 9: A comparison of average localization errors, in meters, and prediction latency across four uncertainty sampling techniques for QuickLoc (QL) as compared to the baseline CNNLOC [26] framework**

## 8.2 Sensitivity Analysis on Device Heterogeneity

Next, we explored the impact of device heterogeneity on achievable latency and localization error for QuickLoc as compared to the non-early exit model in CNNLOC [26].

Each curve in figure 10 depicts the variation in achievable localization error with its associated latency. The curves are captured by varying the threshold parameter of the margin of confidence uncertainty sampling method across both EEA1 and EEA2. The star markings denote the baseline prediction latencies across various devices. We can make two observations from figure 10. First, we note that for the devices excluding GS6, there exist several threshold values in QuickLoc that will produce significant reductions in latency and improved localization error. The reduction in localization error can be attributed to the enhanced learning capabilities introduced by the shorter exit paths that lead to fewer mispredictions due to the vanishing gradient problem. The second observation is that the user can achieve an exponential reduction in localization error by trading off some latency at run-time.

Unfortunately, in this analysis QuickLoc is unable to achieve any improvement in latency for the GS6 device. However, it is important to note that figure 10 only presents results for QuickLoc configurations where both EEA1 and EEA2 are enabled. As we observe from the results in the next subsection, there may be other configurations that deliver better results.

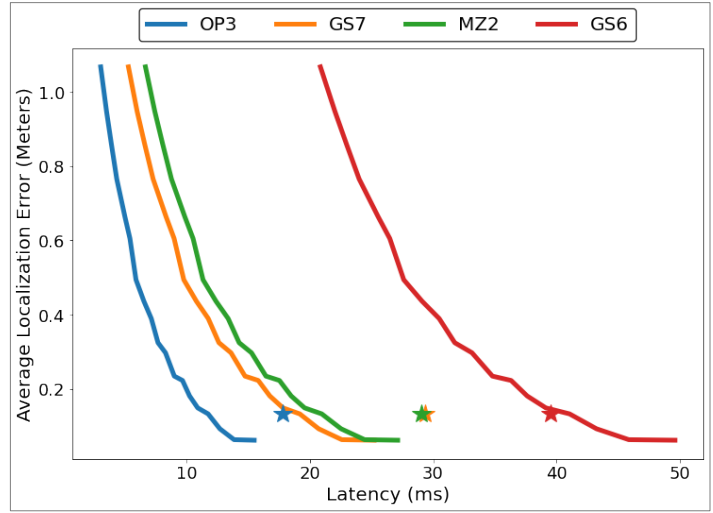

**Figure 10: Achievable localization error with respect to prediction latency for four mobile devices. Baseline localization error and latency for each device is marked by the star symbol (the green and orange stars overlap)**

## 8.3 Analysis of Early Exit Path Configuration

Figure 11 presents the best achievable latency with 95% confidence interval for each mobile device under various early exit configurations while meeting the baseline accuracy target requirements. We found that the best results (least latency) for each device are achieved when EEA1 is disabled (i.e., only EEA2 is enabled) as denoted by the green bars.

In case of the GS6 device, we observe that there are no latency improvements when both the early exit paths are enabled (EEA1+EEA2) compared to CNNLOC. The localization latency further degrades when only EEA1 is enabled for GS6. It is important to note that while different EEA configurations had no impact on the OP3 device, it had a significant impact on the GS6 device. This observation highlights the significance of having multiple EEA paths that can be configured for an unknown device in the online phase.

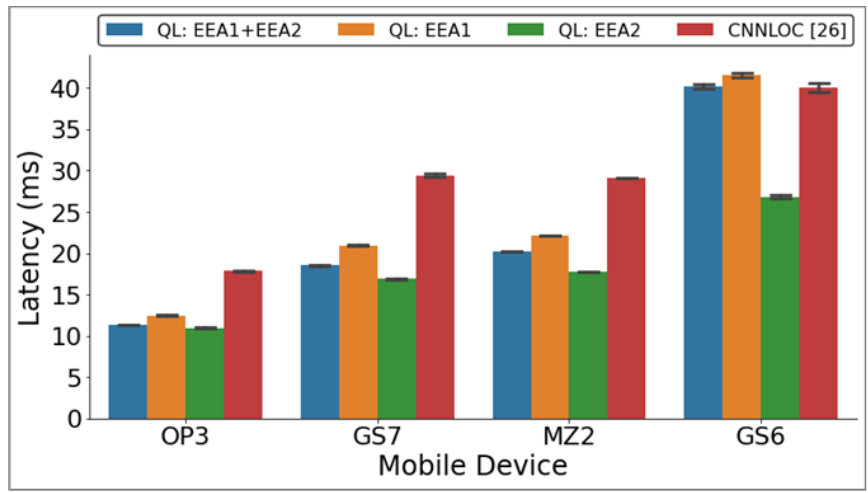

**Figure 11: QuickLoc (QL) device performance under various early exit branch configurations**

## 8.4 Analysis of Inference Energy

The variation in smartphone specifications can greatly impact the energy required to perform a given task. Further, as smartphones are energy constrained devices that run on batteries, prediction latency alone does not dictate framework efficiency. To better highlight the energy savings (energy efficiency) of QuickLoc, we profiled the energy required per location prediction (inference energy) across various smartphones and QuickLoc configurations with 95% confidence interval. The results of this analysis are shown in figure 12.

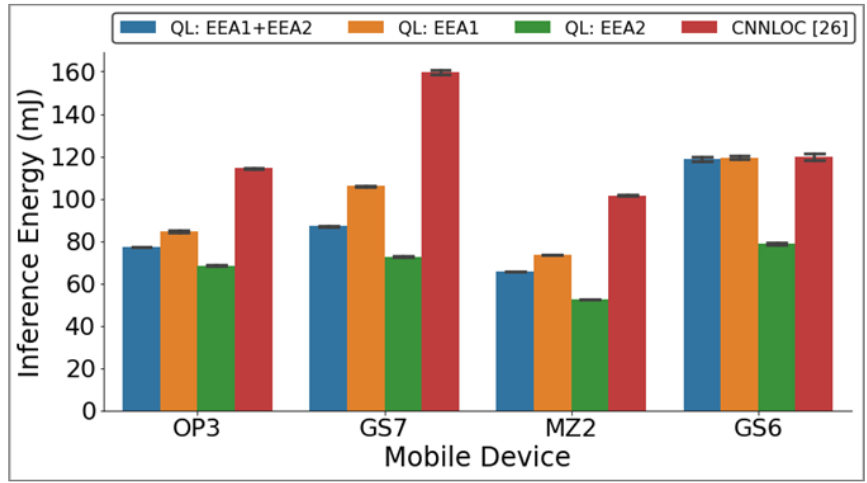

**Figure 12: QuickLoc (QL) inference energy under various early exit branch configurations**

From figure 12, we observe QL: EEA2 consumes the least energy across all devices, as in figure 11. This is because of the large number parameters in the EEA1 output layer that need to be processed every time and are held in memory as the model attempts to exit at EEA1. However, the per-device inference energy trends do not follow the inference latency trends from figure 11. Through figure 12, we observe that the MZ2 device consumes the least energy per prediction as opposed to the OP3 device which has the fastest inference time. In general, we observe up to 45% reduction of inference energy (GS7) with QuickLoc as compared to baseline model.

## 8.5 Analysis on Memory Footprint

In this subsection, we present an analysis of the memory overhead of QuickLoc under various configurations. As the model utilized by QuickLoc has additional layers compared to the baseline work in CNNLOC [26], there is an increase in the memory required to deploy the model on a smartphone.

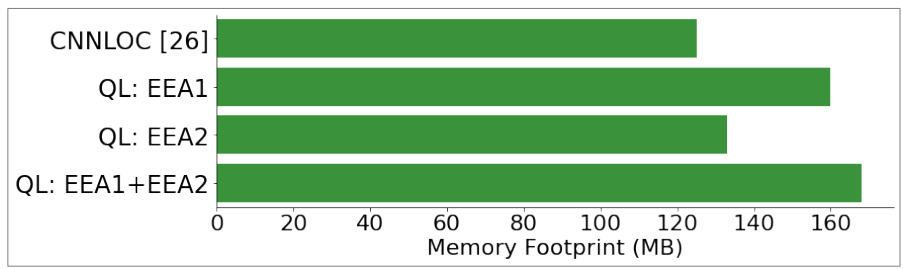

**Figure 13: QuickLoc memory footprint with respect to CNNLOC [26]**

Figure 13 describes the memory footprint of QuickLoc under various early exit configurations as compared to CNNLOC [26]. The most notable observation from figure 13 is the 25% increment in memory footprint when both the early exit branches are enabled (QL: EEA1+EEA2). This is followed by QL: EEA1 which has a 22% increment in memory footprint as compared to CNNLOC [26]. This behavior is mainly attributed to the very large number of parameters in the output later of EEA1 (9.2M parameters) as compared to EEA2 (1.9M parameters). QL: EEA2 only incurs a 3% increase in memory footprint and is therefore the most favorable configuration in general, based on experiments performed in the previous sub-section. From this point onward, we use QL: EEA2 as the default configuration for QuickLoc when comparing it against prior works.

While the memory footprint for QuickLoc is always expected to be higher than the baseline model, the specific increase we observe in our experiments is highly dependent on various factors such as original model complexity, number of early exit paths (or branches) enabled at the time of deployment, and the layer hyperparameters on each early exit path. Due to these factors, we advise caution when adapting ideas from QuickLoc into other model designs.

## 8.6 Analysis on Battery Life

The latency reduction and flexible design of the fingerprinting-based indoor localization model can have a domino effect on various aspects of the overall indoor localization cyber-physical system. One of those aspects is the battery life of a resource constrained smartphone. A longer battery life would translate into the user being able to use their device for real-time localization over a much longer distance. To highlight this point, we have attempted to capture the battery life of various smartphones assuming they were running the localization applications CNNLOC and QuickLoc. These values are then compared to the expected battery life of the smartphone when it is on standby mode with the display turned on.

Table 1 describes the specification of various smartphones under test and includes the battery capacity for each of these devices. We observe that the battery capacity across most devices is close to 3000 mAh.

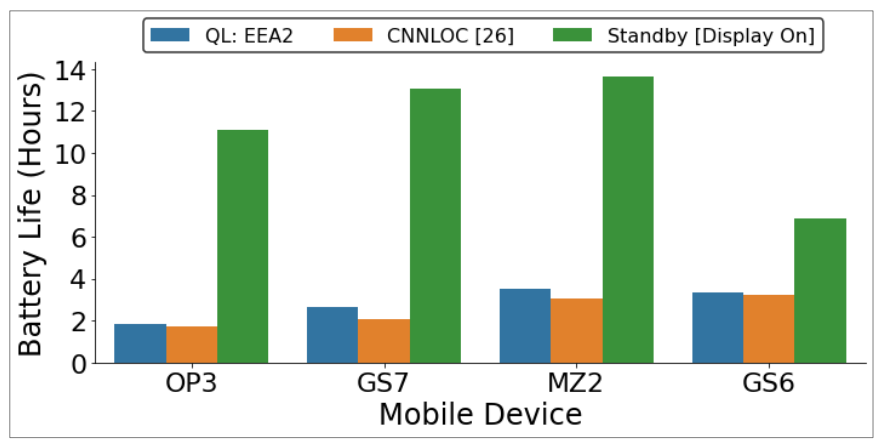

Figure 14: Battery life (hours) for smartphones with QuickLoc (QL: EEA2) and CNNLOC. The smartphones on the horizontal axis are OnePlus 3 (OP3), Samsung Galaxy S7 (GS7), Motorola Z2 (MZ2) and Samsung Galaxy S6 (GS6)

Figure 14 shows the battery life in hours for smartphones under test with QuickLoc and CNNLOC deployed on them, and running continuously while a user performs real-time indoor localization. It can be observed that when running CNNLOC on these smartphones, the battery of the smartphones will last for 1.75 to 3.35 hours, which is an order of magnitude less than the standby battery life of the smartphones. This observation further highlights the importance of optimizing indoor localization frameworks for energy-efficiency. The overall battery life varies across different devices. This can be attributed to the fact that while most devices have very similar battery capacity of 3000 mAh, the current draw across these devices is dependent on factors such as the specific SoC being used, DRAM type and size, the CPU voltage/frequency scheduling strategy that was utilized, etc. Overall, for the smartphones we considered, we observe that the battery life when using deep-learning-based indoor localization frameworks can be increased by up to 28% on smartphones using QuickLoc (GS7). We also noted an increase of about 7%, 15% and 5% for devices OP3, MZ2 and GS6 respectively. We would like to highlight that the results of this analysis are specific to the indoor locales, smartphones, and the deep-model being utilized here. Other settings may yield different improvements.

## 8.7 Overall QuickLoc performance

Figure 15 and 16 describe the accuracy and latency with 95% confidence intervals for QuickLoc (QL: EEA2 variant) as compared to CNNLOC [26], and non-deep machine learning frameworks that employ support vector regression (SVR) [35], and K-nearest Neighbor (KNN) [19]. As we do not cover the impact of device heterogeneity on model accuracy in this work, the results are only presented for the OP3 smartphone. We also utilize the same configuration of QuickLoc (QL: EEA2 with θ_US2=0.8) for both the accuracy and latency results.

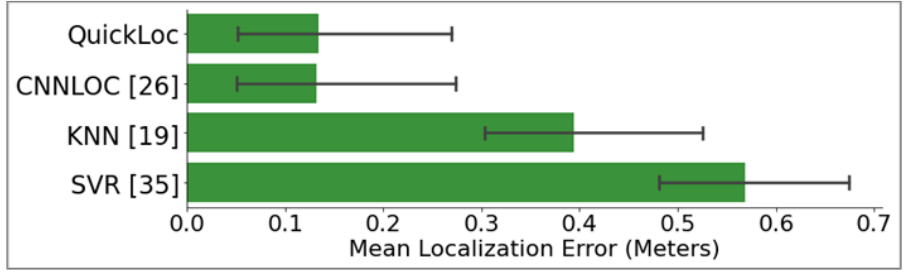

Figure 15: Mean localization error in meters for various indoor localization frameworks.

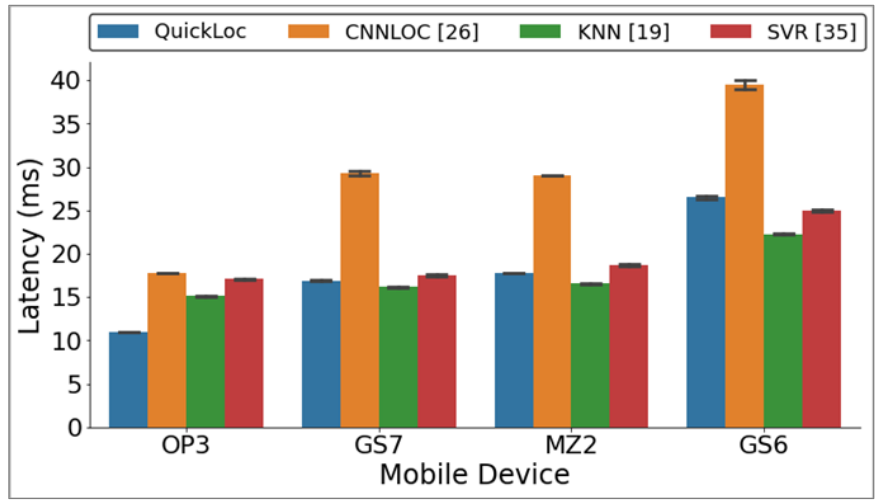

Figure 16: The prediction latency of various indoor localization frameworks with respect to QuickLoc.

From figure 15, we observe that both CNNLOC [26] and QuickLoc deliver a considerable improvement in localization accuracy over KNN [19] and SVR [35]. From the analysis presented in figure 16, we observe that through QuickLoc we are able to achieve up to 42% reduction in prediction latency (GS7) while maintaining our target baseline localization accuracy achieved through CNNLOC [26]. Further, QuickLoc enables us to achieve inference latencies comparable to relatively light-weight non-deep learning indoor localization frameworks in most cases, while outperforming them on the OP3 device. The reason for QuickLoc having lower latency than KNN and SVR on the OP3 device is not entirely clear. We believe that the hardware on the OP3, specifically the DRAM, is geared towards faster and better locality-exploiting burst I/O modes at a cost of higher current draw (1750 mA; in contrast the GS7 only required an average current draw of 1300 mA), which may explain the lower latency for QuickLoc's access patterns on the OP3 device. In summary, the QuickLoc indoor localization framework presented in this work significantly improves prediction latency without any loss in localization accuracy across smartphones and indoor locales. Further, it enables a new form of run-time adaptiveness for deep-learning-based indoor localization frameworks that trades-off localization accuracy, inference latency, and energy against run-time memory footprint.

## 9 GENERALITY OF PROPOSED APPROACH

In this section, we highlight the generality and the versatile nature of our proposed approach by applying it to two other deep learning-based indoor localization frameworks proposed in [64] and [65]. We first present a discussion of the proposed works in [64] and [65] and our changes to their model. Later, we use Wi-Fi fingerprints from our own dataset in section 7.2 to train the model in [64] and the dataset in [24] to train the model in [65]. We perform a brief sensitivity analysis on uncertainty threshold for the early exit variations of these models and later compare their prediction accuracies and latencies with their baseline counterparts.

### 9.1 Depthwise Separable Convolutions Based Network

The work in [64] utilizes depthwise separable convolutions for passive indoor localization (DSCP). It is architecturally similar to MobileNets. Unfortunately, this work employs Channel State Information (CSI) as location fingerprints and this dataset is not publicly available. To overcome this issue, we train a model based on the same concepts as in [64] and with the RSSI-based dataset used in this work. Figure 17 depicts the design of the baseline DSCP inspired model along with the early exit path. The DSCP model mainly consists of a convolutional layer followed by pairs of depthwise and pointwise convolutional layers. Considering that the model in figure 17 is designed to work with the dataset from this work (section 7.2), the output layer consists of 342 neurons. These neurons in the output layer are representative of every localizable location (reference points) across all paths discussed in section 7.2 (similar to QuickLoc). Based on the initial intuition behind our work, we found that a shallow neural network path delivers the majority of accurate results. For brevity figure 17 includes the most optimal early exit path that was chosen after various design iterations.

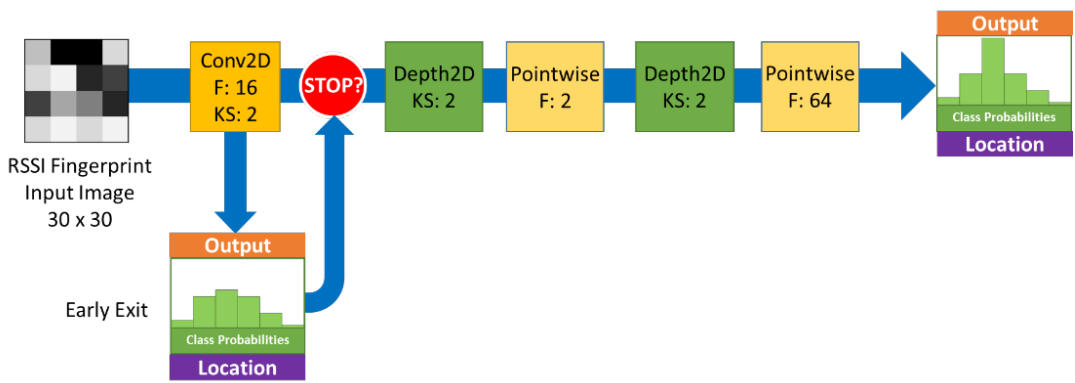

Figure 17: DSCP [64] inspired convolutional model along with early exit path for indoor localization

To identify a suitable value of uncertainty threshold, we performed a sensitivity analysis as in section 8.1. In figure 18, the left and the right vertical axes represent the mean localization error (red) and model latency (green) respectively, captured using the Oneplus3 smartphone. The horizontal axis represents the entropy-based uncertainty threshold varied in the range of 0.01 to 0.50 with a step size of 0.02. The dotted lines represent the mean localization error and model latency for the baseline DSCP model without the early exit path, while the solid lines capture these metrics for the early exit DSCP (EE-DSCP) model. With an aim of keeping localization error under 1 meter, we observe that the localization latency can be reduced by 80% (2ms) for a threshold value of 0.36.

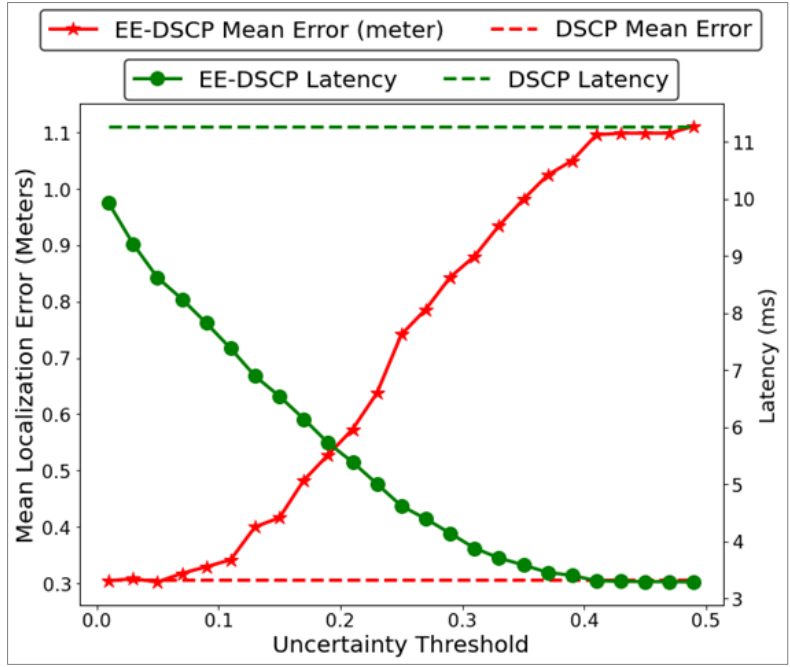

**Figure 18: Sensitivity of mean localization error and latency towards entropy-based uncertainty threshold**

Based on our observations from figure 18, we choose the uncertainty threshold value of 0.03 as it maintains the baseline localization performance. Figure 19 captures the mean inference latency and mean localization error for the baseline DSCP model (no early exit) and the early exit DSCP model (EE-DSCP) as observed for the Oneplus 3 smartphone. From figure 19(a), we observed that for the value of 0.03 delivers ~25% reduction (11ms to 8ms) in localization latency with a slight improvement in accuracy as observed in figure 19(b).

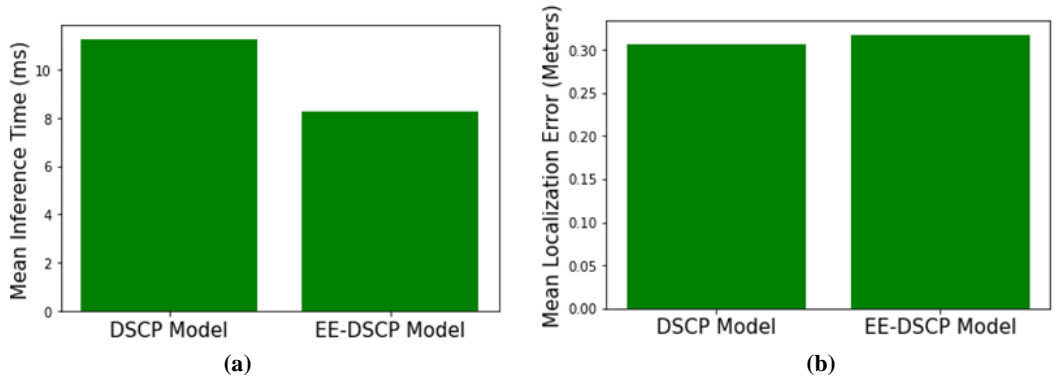

**Figure 19: The mean inference times and localization errors of the EE-DSCP model with respect to baseline DSCP model**

## 9.2 Predicting Buildings and Floors for the UJIndoorLoc Dataset

The work by Jang et al., [65], presents a convolutional neural network model utilized to predict the building and floor for the RSSI fingerprint dataset UJIndoorLoc given in [24]. The UJIndoorLoc dataset [24] consists of three buildings: The first two buildings consist of four floors each and the third building contains five floors. This produces a total of 13 combinations of floors and buildings. The model in [65] attempts to predict the floor (and building) the user is on given the WiFi RSSI fingerprint. More information on the dataset and model can be found in [24] and [65]. We chose this work as another deep learning indoor localization framework to showcase the generalizability of QuickLoc. The convolutional model as presented in [65] with the addition of an early exit path, based on our proposed approach, is depicted in figure 20. It employs convolutional layers and down-sampling through pooling. It additionally consists of fully connected layers before output layers. The output layers in figure 20 consist 13 neurons (13 output classes), each corresponding to a unique building and floor.

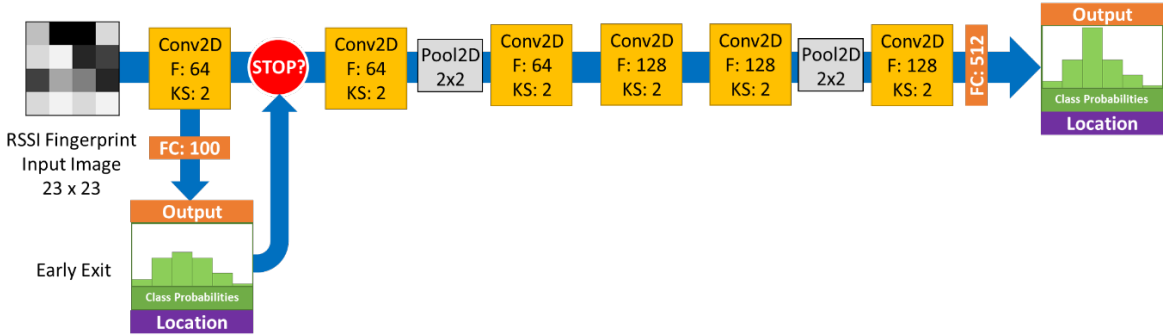

**Figure 20: The convolutional model as presented in [2] with the addition of an early exit**

Similar to the previous implementation (section 9.1), we performed a sensitivity analysis on the entropy-based uncertainty threshold with respect to accuracy and latency of the model in [65]. The observations of this analysis are presented in figure 21. It is important to note that for this analysis we employ top-1 accuracy instead of mean localization error. We do so because the work in [65] is designed to predict the building and floors instead of specific position within a floorplan. Therefore, we cannot conventionally compute localization error in this case.

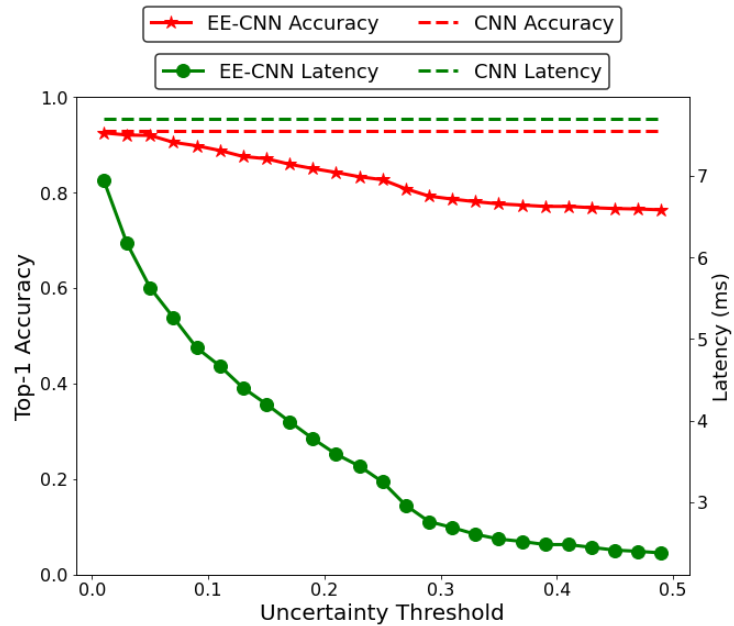

**Figure 21: Sensitivity of accuracy and latency towards entropy-based uncertainty threshold for model in [65] on a OnePlus 3 smartphone**

From figure 21, we observe that as the value of uncertainty threshold is increased the latency and accuracy of the model generally drop. However, for small values of the threshold, the accuracy remains the same while the latency drops considerably. We chose the value of 0.03 for uncertainty threshold as it maintains the similar levels of accuracy as the baseline CNN model, but reduces latency by 20% as depicted in figure 22 for the OnePlus 3 device. We use the same dataset as in [65].

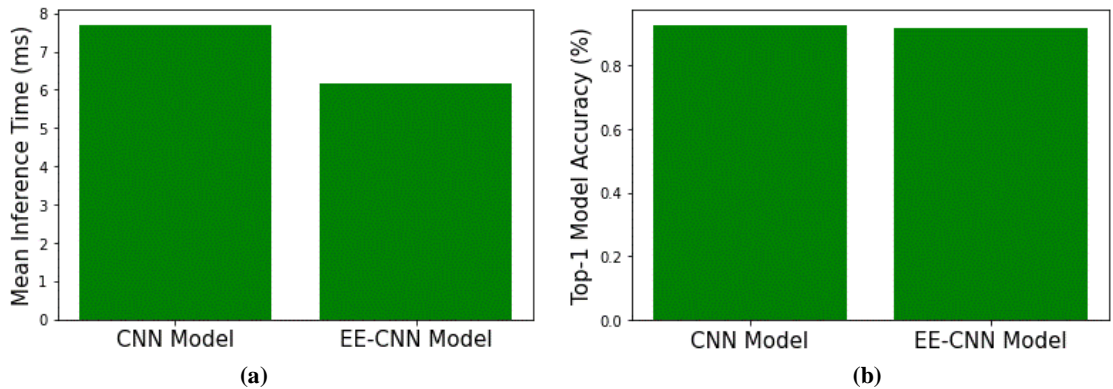

**Figure 22: The mean inference times and model accuracy of the CNN model in [65] with respect to its EE-CNN**

In summary, through the implementation of our approach on the works in [64] and [65], we have demonstrated that our approach with QuickLoc can be easily generalized to other deep-learning based indoor localization frameworks.

# 10  CONCLUSIONS AND FUTURE WORK

In this paper, we presented an in-depth analysis of a deep learning based indoor localization framework that is expected to deliver accurate results on various mobile devices in real-time. Our analysis highlighted the significant lack of consistent performance across varying deep learning model depths and across diverse mobile devices. To overcome this challenge, we proposed the novel QuickLoc framework, that is able to adapt the localization latency for the target device through early exit strategies and reduce average localization error at the same time.

As a part of future work, we will be focusing on further improving the quality of localization. Towards this goal, a possible extension of our work could be to explore the impact of memory latency and bandwidth on prediction latencies. It may also be interesting to explore QuickLoc's behavior for hybrid indoor localization strategies that utilize deep learning models in conjunction with other approaches such as particle filters, dead reckoning, and Markov State Machines. In such cases, early-exit attempts could be tuned in response to information from other approaches.

### ACKNOWLEDGMENTS
This research was supported by the National Science Foundation (NSF) under grant number ECCS-1646562

### REFERENCES
[9] Richard B. Langley, "The evolution of the GPS receiver," GPS World, vol. 11, no. 4, pp. 54-58, 2000
[10] "A brief history of GPS," 2019 [Online]. Available: https://www.pcworld.com/article/2000276/a-brief-history-of-gps
[11] "Wi-Fi RTT (IEEE 802.11mc)," 2019 [online]. Available: https://www.source.android.com/devices/tech/connect/wifi-rtt
[12] "Top 33 indoor localization services in the US," 2019 [online]. Available: https://www.technavio.com/blog/top-33-indoor-location-based-services-bs-companies-in-the-us
[13] Bruno Raffaele, and Franca Delmastro, "Design and analysis of a bluetooth-based indoor localization system," International Conference on Personal Wireless Communications, 2003
[14] Faheem Zafari, Athanasios Gkelias, and Kin K. Leung, "A survey of indoor localization systems and technologies," Communications Surveys & Tutorials, vol. 21, no. 3, pp. 2568-2599, 2019
[15] Krishna Chintalapudi, Anand Padmanabha Iyer, Venkata N. Padmanabhan, "Indoor localization

[15] without the pain," International Conference on Mobile Computing and Networking (MobiCom), 2010
[16] Andreas Haeberlen, Eliot Flannery, Andrew. M. Ladd, Algis Rudys, Dan S. Wallach, and Lydia E. Kavraki, "Practical robust localization over large-scale 802.11 wireless networks," International Conference on Mobile Computing and Networking (MobiCom), 2004
[17] Prajindra Krishnan, A. Krishnakumar, Wen-Hu Ju, Collin Mallows, and Sachine Gamt, "A system for LEASE: Location estimation assisted by stationary emitters for indoor RF wireless networks," International Conference on Computer Communications (INFOCOM), 2004
[18] Jie Xiong, and Kyle Jamieson, "Towards fine-grained radio-based indoor location," Mobile Computing Systems & Applications (HotMobile), 2012
[19] Elahe Soltanaghaei, Avinash Kalyanaraman, and Kamin Whitehouse, "Multipath Triangulation: Decimeter-level WiFi Localization and Orientation with a Single Unaided Receiver," Mobile Systems, Applications, and Services (MobiSys), 2018
[20] Abdulrehman Alarifi, Abdulmalik Al-Salman, Mansour Alsaleh, Ahmad Alnafessah, Suheer Al-Hadhrami, Mai A. Al-Ammar, and Hend S. Al-Khalifa, "Ultra wideband indoor positioning technologies: Analysis and recent advances," Sensors, vol. 16, no. 5, pp. 707, 2016
[21] Patrick Dickinson, Gregorz Cielniak, Olivier Szymanezyk, and Mike Mannion, "Indoor positioning of shoppers using a network of Bluetooth Low Energy beacons," Indoor Positioning and Indoor Navigation, 2016
[22] Seng-Yong. Lau, Tsung-Han Lin, Te-Yuan Huang, I-Hei Ng, and Polly Huang, "A measurement study of zigbee-based indoor localization systems under RF interference," Workshop on Experimental evaluation and Characterization (WIN-TECH), 2009
[23] Ugur Bolat, and Mehmat Akcakoca, "A hybrid indoor positioning solution based on Wi-Fi, magnetic field, and inertial navigation," Workshop on Positioning, Navigation and Communications (WPNC), 2017
[24] Joaquín Torres-Sospedra, Raúl Montoliu, Adolfo Martínez-Usó, Joan P. Avariento, Tomas J. Arnau, Mauri Benedito-Bordonau, and Joaquín Huerta, "UJIIndoorLoc: A new multi-building and multi-floor database for WLAN fingerprint-based indoor localization problems," Indoor Positioning and Indoor Navigation (IPIN), 2014
[25] Paramvir Bahl, and Venkata. Padmanabhan, "RADAR: An in-building RF-based user location and tracking system," International Conference on Computer Communications (INFOCOM), 2000
[26] "IndoorAtlas, Yahoo team geomagnetic building mapping in Japan," 2016 [Online] Available: https://www.mediapost.com/publications/article/269899/indooratlas-yahoo-team-geomagnetic-building-mappi.html
[27] Sudeep Pasricha, Vinay Ugave, Charles W. Anderson, and Qi Han. "LearnLoc: A Framework for Smart Indoor Localization with Embedded Mobile Devices," International Conference on Hardware/Software Codesign and System Synthesis (CODES+ ISSS), 2015
[28] Wei Zhang, Kan Liu, Weidong Zhang, Youmei Zhang, and Jason Gu, "Deep Neural Networks for wireless localization in indoor and outdoor environments," Neurocomputing, vol. 194, pp. 279-287, 2016
[29] Xuyu Wang, Lingjun Gao, Shiwen Mao, and Santosh Pandey, "DeepFi: Deep learning for indoor fingerprinting using channel state information," Wireless Communications and Networking Conference (WCNC), 2015
[30] Hao Chen, Yifan Zhang, Wei Li, Xiafeng Tao and Ping Zhang, "ConFi: Convolutional Neural Networks Based Indoor Wi-Fi Localization Using Channel State Information," IEEE Access, vol. 5, pp. 18066-18074, 2017
[31] Hong Jiang, Chao Peng and Jing Sun, "Deep Belief Network for Fingerprinting-Based RFID Indoor Localization," International Conference on Communications (ICC), 2019
[32] Priyadarshini Panda, Abhronil Sengupta, and Kaushik Roy, "Conditional deep learning for energy-efficient and enhanced pattern recognition," Design, Automation & Test in Europe Conference & Exhibition, 2016
[33] Surat Teerapittayanon, Brandley McDanel, and Hsiang-Tsung Kung, "Branchynet: Fast inference via early exiting from deep neural networks," International Conference on Pattern Recognition (ICPR), 2016
[34] Ayush Mittal, Saideep Tiku, and Sudeep Pasricha, "Adapting Convolutional Neural Networks for Indoor Localization with Smart Mobile Devices," Great Lakes Symposium on VLSI (GLSVLSI), 2018
[35] Yann LeCun, Leon Bottou, Yoshua Bengio, and Patrick Haffner, "Gradient-based learning applied to

document recognition," Proceedings of the IEEE, vol. 86, no. 11, pp. 2278-2324, 1998
[36] Alex Krizhevsky, IIya Sutskever, and Geoffrey E, Hinton, "ImageNet classification with deep convolutional neural networks," Advances in neural information processing systems (NIPS), 2012
[37] Robert Paróczai, and László Kocsis, "Analysis of human walking and running parameters as a function of speed," Technology and Health Care, vol. 14, no. 4, pp. 251-260, 2006
[38] Claude Elwood Shannon, "A mathematical theory of communication," Bell System Technical Journal, vol. 27, no. 3, pp.379-423, 1948
[39] Gao Huang, Zhuang Liu, Laurens Maaten, and Killian Q. Weinberger, "Densely connected convolutional networks," Conference on Computer Vision and Pattern Recognition (CVPR), 2017
[40] Sungwon Yang, Pralav Dessai, Mansi Verma and Mario Gerla, "FreeLoc: Calibration-free crowdsourced indoor localization," International Conference on Computer Communications (INFOCOM), 2013
[41] "Packet Capture," 2020 [Online]. Available: https://play.google.com/store/apps/details?id=jp.co.taosoftware.android.packetcapture
[42] "Nexmon Driver Patching," 2020 [Online]. Available: https://github.com/seemoo-lab/nexmon
[43] Yen-Kai Cheng, Hsin-Jui Chou, and Ronald Y. Chang, "Machine-Learning Indoor Localization with Access Point Selection and Signal Strength Reconstruction", Vehicular Technology Conference (VTC), 2016
[44] "Tensorflow Benchmark Performance," 2020 [Online]. Available: https://www.tensorflow.org/lite/performance/benchmarks
[45] "Battery Manager API," 2020 [Online]. Available: https://developer.android.com/reference/android/os/BatteryManager
[46] Andrew Mackey, Petros Spachos, Liang Song and Konstanitos. N. Plataniotis, "Improving BLE Beacon Proximity Estimation Accuracy Through Bayesian Filtering," Internet of Things Journal, vol. 7, no. 4, pp. 3160-3169, 2020
[47] "Fraunhofer IIS uses Awiloc indoor positioning magic to guide museum patrons," 2020 [online]. Available: https://www.engadget.com/2010-12-13-fraunhofer-iis-uses-awiloc-indoor-positioning-magic-to-guide-mus.html
[48] Wenhua Shao, Haiyong Luo, Fang Zhao, Yan Ma, Zhongliang Zhao, and Antonino Crivello, "Indoor Positioning Based on Fingerprint-Image and Deep Learning," IEEE Access, vol. 6, pp. 74699-74712, 2018
[49] Zhuan Gu, Zeqin Chen, Yuexing Zhang, Ying Zhu, MingMing Lu, Ai Chen, "Reducing fingerprint collection for indoor localization," Computer Communications, vol. 83, pp. 56-63, 2016
[50] Vahideh Moghtadaiee, Seyed Ali Ghorashi and Mohammad Ghavami, "New Reconstructed Database for Cost Reduction in Indoor Fingerprinting Localization," IEEE Access, vol. 7, pp. 104462-104477, 2019
[51] Yuexing Zhang, Ying Zhu, Mingming Lu and Ai Chen, "Using compressive sensing to reduce fingerprint collection for indoor localization," IEEE Wireless Communications and Networking Conference (WCNC), 2013
[52] Han Zou et al., "Adversarial Learning-Enabled Automatic WiFi Indoor Radio Map Construction and Adaptation With Mobile Robot," in IEEE Internet of Things Journal, vol. 7, no. 8, pp. 6946-6954, 2020
[53] Saideep Tiku, Sudeep Pasricha, Branislav Notaros, and Qi Han, "SHERPA: A Lightweight Smartphone Heterogeneity Resilient Portable Indoor Localization Framework," International Conference on Embedded Software and Systems (ICESS), 2019
[54] Han Zou, Baoqi Huang, Xiaoxuan Lu, Hao Jiang, and Lihua Xie, "Standardizing location fingerprints across heterogeneous mobile devices for indoor localization," Wireless Communications and Networking Conference (WNC), 2016
[55] Mikkel Baun Kjærgaard, "Indoor location fingerprinting with heterogeneous clients," Pervasive and Mobile Computing, vol. 7, no. 1, pp. 31-43, 2010
[56] Yan Li, Simon Williams, Bill Moran and Allison Kealy, "A Probabilistic Indoor Localization System for Heterogeneous Devices," IEEE Sensors Journal, vol. 19, no. 16, pp. 6822-6832, 2019
[57] A.K.M. Mahtab Hossain, Yunye Jin, Wee-Seng Soh and Hien Nguyen Van, "SSD: A Robust RF Location Fingerprint Addressing Mobile Devices' Heterogeneity," IEEE Transactions on Mobile Computing, vol. 12, no. 1, pp. 65-77, 2013
[58] "Enhance the Educational Experience with Wayfinding and Other Location-Based Services," 2020 [online]. Available: http://www.bluepath.me/use-cases-indoor-navigation/educational.php

[59] "You Safe: The App for Emergency Evacuations developed by DB System with Situm's indoor location technology," 2020 [online]. Available: https://situm.com/en/success-stories/you-safe-the-app-for-emergency-evacuations-developed-by-db-system-with-situms-indoor-location-technology/

[60] Itay Hubara, Matthieu Courbariaux, Daniel Soudry, Ran El-Yaniv, and Yoshua Bengio, "Quantized neural networks: training neural networks with low precision weights and activations," The Journal of Machine Learning Research, vol. 18, no. 1, pp. 6869–6898, 2017

[61] Tien-Ju Yang, Yu-Hsin Chen, and Vivienne Sze, "Designing Energy-Efficient Convolutional Neural Networks Using Energy-Aware Pruning," Conference on Computer Vision and Pattern Recognition (CVPR), 2017

[62] Hao Ni, Jie Shen, and Chong Yuan, "Enhanced Knowledge Distillation for Face Recognition," IEEE Parallel & Distributed Processing with Applications, Big Data & Cloud Computing, Sustainable Computing & Communications, 2019

[63] Yuge Zhang, et al., "Deeper insights into weight sharing in neural architecture search." arXiv preprint arXiv:2001.01431, 2020

[64] W. Xun, L. Sun, C. Han, Z. Lin and J. Guo, "Depthwise Separable Convolution based Passive Indoor Localization using CSI Fingerprint," Wireless Communications and Networking Conference (WCNC), 2020

[65] J. Jang and S. Hong, "Indoor Localization with WiFi Fingerprinting Using Convolutional Neural Network," International Conference on Ubiquitous and Future Networks (ICUFN), 2018